\documentclass[preprint,aps,showpacs]{revtex4}

\usepackage{graphicx}
\usepackage{bm}       
\usepackage{mathptmx} 

\begin{document}

\title{One flavor QCD}
\author{Michael Creutz}
\affiliation{
Physics Department, Brookhaven National Laboratory\\
Upton, NY 11973, USA
}

\begin{abstract}
One flavor QCD is a rather intriguing variation on the underlying
theory of hadrons.  In this case quantum anomalies remove all chiral
symmetries.  This paper discusses the qualitative behavior of this
theory as a function of its basic parameters, exploring the
non-trivial phase structure expected as these parameters are varied.
Comments are made on the expected changes to this structure if the
gauge group is made larger and the fermions are put into higher
representations.
\end{abstract}
\maketitle

\section{Introduction}

QCD, the non-Abelian gauge theory of quarks and gluons, is now
generally regarded as the underlying dynamics of strongly interacting
hadrons.  It is a very economical theory, with the only parameters
being the overall scale, usually called $\Lambda_{qcd}$, and the quark
masses.  Indeed, in the zero quark mass limit we have a theory with
essentially no parameters, in that all dimensionless ratios should in
principle be determined.  With several massless quarks this theory has
massless Goldstone bosons representing a spontaneous breaking of
chiral symmetry.  The fact that the pions are considerably lighter
than other hadrons is generally regarded as a consequence of the quark
masses being rather small.

The one flavor situation, while not phenomenologically particularly
relevant, is fascinating in its own right.  In this case quantum
mechanical anomalies remove all chiral symmetries from the problem.
No massless Goldstone bosons are expected, and there is nothing to
protect the quark mass from additive renormalization.  It appears
possible to have a massless particle by tuning the quark mass
appropriately, but this tuning is not protected by any symmetry and
occurs at a mass value shifted from zero.  The amount of this shift is
non-perturbative and scheme dependent.

\begin{figure*}
\centering
\includegraphics[width=3in]{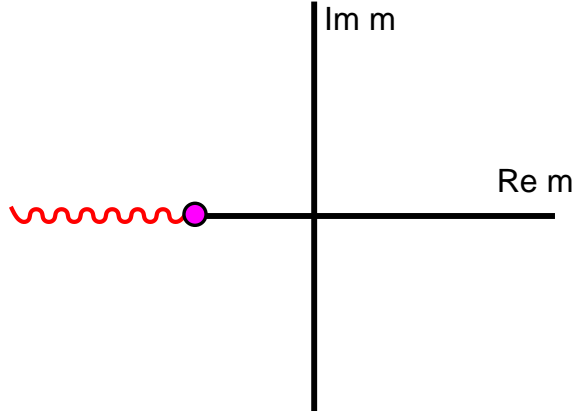}
\caption{ The phase diagram for one flavor quark-gluon dynamics in the
 complex mass plane.  The wavy line in the negative mass region
 corresponds to a first order phase transition ending at a second
 order critical point.  The transition represents a spontaneous
 breaking of CP symmetry.  A finite gap separates the transition from
 vanishing quark mass.}
\label{oneflavor} 
\end{figure*}

In the one flavor theory, the classical formulation does have a chiral
symmetry in parameter space.  If the mass is complexified in the sense
described in the next section, physics is naively independent of the
phase of this parameter.  However, when quantum effects are taken into
account, a non-trivial dependence on this phase survives and a
rather interesting phase diagram in complex mass appears.  A large
negative mass should be accompanied by a spontaneous breakdown of
parity and charge conjugation symmetry, as sketched in
Fig.~(\ref{oneflavor}).  

Despite the simplicity of this diagram, certain aspects of this theory
remain controversial.  Does chiral symmetry have anything to say about
the massless theory?  What does one really mean by a massless quark
when it is confined?  Is there any sense that a quark condensate can
be defined?  Is it in any sense an ``order parameter''?  The purpose
of this paper is to provide a framework for discussing these issues by
bringing together a variety of arguments that support the basic
structure indicated in Fig.~(\ref{oneflavor}).

After a brief discussion of the parameters of the theory in Section
\ref{parameters}, Section \ref{effective} shows how simple effective
Lagrangian arguments give the basic structure.  Much of that section
is adapted from Ref.~\cite{Creutz:2000bs}.  A discussion of the Dirac
eigenvalue structure for this theory appears in sections
\ref{eigenvalues} and \ref{zeromodes}.  Small complex eigenvalues are
treated separately from the exact zero modes arising from
topologically non-trivial gauge fields.  These sections expand on the
ideas in Ref.~\cite{Creutz:2005rd}.  Section \ref{rgroup} expands on
Ref.~\cite{Creutz:2004fi}, reviewing the renormalization group
arguments for regularizing the theory, showing how the renormalized
mass is defined, and exposing the ambiguity in this definition for the
one flavor theory.  Section \ref{colors} addresses expected changes in
the basic picture when the size of the gauge group is increased and
the fermions placed in higher representations than the fundamental.
Brief conclusions are in Section \ref{conclude}.

\section{Parameters}
\label{parameters}
The theory under consideration is motivated by the classical
Lagrangian density for $SU(3)$ gauge fields coupled to one species of
quark in the fundamental representation of the gauge group
\begin{equation}
L={1\over 4 g^2} F_{\mu\nu}F_{\mu\nu}
+\overline\psi \gamma_\mu\partial_\mu \psi
-m \overline\psi\psi.
\end{equation}
Here the color indices associated with the gauge symmetry are
suppressed.  Of course we are really interested in the quantum theory,
which requires definition with an ultraviolet regulator, such as the
lattice.  In this sense the coupling $g$ and the mass $m$ should be
regarded as bare parameters.  As the cutoff is removed, the bare
parameters go to zero, as described by the simple renormalization
group analysis reviewed in Section \ref{rgroup}.

In this Lagrangian the change of variables
\begin{equation}
\matrix{
&\psi \rightarrow e^{i\gamma_5\theta/2}\psi\cr
&\overline\psi \rightarrow
\overline\psi e^{i\gamma_5\theta/2}\cr
}
\label{chiralrot}
\end{equation}
leaves the kinetic term for the fermions unchanged, but does not leave
the mass term invariant
\begin{equation}
m \overline\psi\psi\rightarrow m\cos(\theta)\overline\psi \psi
+ im\sin(\theta)\overline\psi \gamma_5\psi
=m_1\overline\psi \psi+m_5\overline\psi \gamma_5\psi.
\end{equation}
So, if we allow a CP violating mass term of form
$i\overline\psi\gamma_5\psi$, the theory appears to depend on three
parameters $g$, $m$ and $\theta$ or equivalently $g$, $m_1$ and $m_5$.

Introducing left and right fields
\begin{equation}
\psi_{R,L}={1\over 2}(1\pm \gamma_5)\psi
\end{equation}
the rotated mass term is usually rewritten in terms of a
complexification of the quark mass
\begin{equation}
{\rm Re}\ m \overline\psi \psi
+ i\ {\rm Im}\ m\sin(\theta)\overline\psi \gamma_5\psi
= m\ \overline\psi_L\psi_R + m^*\ \overline\psi_R\psi_L.
\end{equation}
Then the rotation of Eq.~(\ref{chiralrot}) corresponds to a change in
the phase of $m$.

As we are just doing a change of variables, the physics of the
classical theory should be independent of the phase of $m$.  However,
quantum mechanically this symmetry fails due to the non-invariance of
the path integral measure.  This is not entirely obvious since the
determinant of the transformation factor $e^{i\gamma_5\theta/2}$ would
appear to be unity since $\gamma_5$ is traceless.  However all known
regulators break this symmetry.  How this physics manifests itself in
a particular theoretical formalism depends on approach.  With a
Pauli-Villars regulator \cite{Pauli:1949zm}, the heavy auxiliary field
has a phase in its mass; the theta parameter is the relative phase
between the regulator mass and the physical fermion mass.  In other
continuum schemes the phase is often pushed into the path integral
measure \cite{Fujikawa:1979ay}.  With Wilson fermions we have the
added Wilson term, which is itself of a mass like form.  Theta then
becomes a relative phase between the Wilson term and the explicit mass
\cite{Seiler:1981jf}.  With overlap fermions the chiral rotation
involves two different versions of $\gamma_5$, one of which is not
always traceless.

On renormalization, discussed in more detail in Section \ref{rgroup},
these three parameters, $\{g,\ {\rm Re}\ m,\ {\rm Im}\ m\}$, are
replaced by the overall scale of the strong interactions, usually
called $\Lambda_{qcd}$, and renormalized values for the real and
imaginary parts of the quark mass.  Fig.~(\ref{oneflavor}) represents
the resulting phase diagram in the latter two parameters.

Many discussions in the literature trade the imaginary part of the
mass for a topological term $\tilde F_{\mu\nu} F_{\mu\nu}$ in the
Lagrangian.  This quantity is a total divergence, and its integral
over all space-time is proportional to an integer winding number.  The
angle conjugate to this integer might be thought of as a further
parameter.  However, this variable and the phase of the mass are
actually redundant since the anomaly allows one to rotate between
them.  Thus there is only one physically significant angle on which
the theory depends.  For this paper we adopt the convention that any
topological term in the Lagrangian has been rotated into the phase of
the mass.  

After this rotation, traditional discussions use the magnitude and the
phase of the mass as independent variables.  However, we will see that
because of ambiguities in defining the quark mass, the magnitude and
phase are potentially singular coordinates.  Thus it is cleaner to use
as parameters the real and imaginary parts of the renormalized mass.
This is an issue special to the one flavor theory; with multiple
degenerate fermions, broken chiral symmetry with its attendant
Goldstone bosons uniquely defines the massless theory.

\section{Effective Lagrangians}
\label{effective}

Outside of the lattice, effective chiral Lagrangians have long been
among our most powerful tools to investigate non-perturbative
phenomena.  They build in the known chiral symmetries and have been
highly successful in describing the physics of the light pseudoscalar
mesons.  Hints of the possibility for spontaneous CP violation in this
approach go back some time \cite{dashen}.  Extensions to study the
behavior with complex mass appear in several references
\cite{DiVecchia:1980ve,Leutwyler:1992yt,Smilga:1998dh}.  In this
section I will rehash some of these arguments in the context of the
one flavor theory.

A quick but imprecise argument gives the expected picture.  With only
one flavor, there is only one light pseudo-scalar meson, referred to
here as the $\eta^\prime$.  Were it not for anomalies, conventional
chiral symmetry arguments suggest the mass squared of this particle
would go to zero linearly with the quark mass,
\begin{equation}
m_{\eta^\prime}^2 \sim m_q.
\end{equation}
Throughout this paper we assume that appropriate powers of the strong
interaction scale, $\Lambda_{qcd}$ are inserted for dimensional
purposes.  But, just as the $\eta^\prime$ in the three flavor theory
gets mass from anomalies, a similar contribution should appear here;
assume it is simply an additive constant
\begin{equation}
m_{\eta^\prime}^2 \sim m_q+c.
\label{shift}
\end{equation}
Consider describing this model by an effective potential for the
$\eta^\prime$ field.  This should include the possibility for these particles
to self interact, suggesting something like
\begin{equation}
V(\eta^\prime)= {m_q+c\over 2} {\eta^\prime}^2 + \lambda {\eta^\prime}^4.
\end{equation} 
To get the phase diagram of Fig.~(\ref{oneflavor}), we add in a $m_5$
term as a linear piece in the eta prime field
\begin{equation}
V(\eta^\prime; m_1,m_5)= {m_1+c\over 2} {\eta^\prime}^2 
+ \lambda {\eta^\prime}^4 + m_5 \eta^\prime.
\label{elone}
\end{equation} 
At $m_1 < -c$ the effective mass of the eta prime goes negative.  This
will give a spontaneous breaking in the canonical manner, with the
field acquiring an expectation value
\begin{equation}
\langle \eta^\prime \rangle \sim \langle \overline\psi
\gamma_5\psi\rangle \sim \sqrt
{
|m_1|-c
\over 
4\lambda
}
\ne 0.
\end{equation}
As this is a CP odd field, CP is spontaneously broken.  In particular,
vertices connecting odd numbers of physical mesons will not vanish,
unlike in the unbroken theory where the number of pseudoscalar mesons
is preserved modulo 2.

Note that this transition occurs at a negative quark mass, and nothing
special happens at $m_1=0$.  Of course the bare quark mass is not
really physical since it is a divergent quantity in need of
renormalization.  Normally in the multiple flavor theory chiral
symmetry forces this renormalization to be multiplicative, making
vanishing mass special.  However with only one flavor there is no
chiral symmetry; thus, there is nothing to prevent an additive shift
in this parameter.  We will see later how such a shift is generated
non-perturbatively by topologically non-trivial gauge configurations.
Despite this, with a cutoff in place, these qualitative arguments
suggest it is only for a negative quark mass that this parity
violating phase transition will take place.  And this also suggests
that the magnitude of this gap is of order the square of the eta prime
mass.  As the latter is of order the strong interaction scale, the
critical mass for the CP violating transition is of order
$\Lambda_{qcd}$, as it must be.

This argument is suggestive but certainly not rigorous.  To lend more
credence to this qualitative picture, note that a similar phenomenon
occurs in two dimensional electrodynamics.  The Schwinger model is
exactly solvable at zero bare mass, with the spectrum being a free
massive boson.  However, for negative bare mass, qualitative
semi-classical arguments indicate the same structure as discussed
above, with a spontaneous generation of a parity violating background
electric field.  Under the bosonization process
\cite{Coleman:1975pw,Coleman:1976uz}, the quark mass term corresponds
to a sinusoidal term in the effective potential for the scalar field
\begin{equation}
m\overline\psi \psi \leftrightarrow \xi m\cos(2\sqrt\pi\eta^\prime)
\end{equation}
where $\xi$ is a numerical constant.  Regularization and normal
ordering are required for a proper definition but are not important
here.  Combining this with the photon mass from the anomaly suggests
an effective potential for the $\eta^\prime$ field of form
\begin{equation}
V(\eta^\prime) \sim {e^2\over 2\pi} {\eta^\prime}^2 
- \xi m \cos(2\sqrt\pi\eta^\prime).
\end{equation}
For small positive $m$, the second term introduces multiple meson
couplings, making the theory no longer free.  It also shifts the boson
mass
\begin{equation}
m_{\eta^\prime}^2 \sim {e^2\over \pi}+4\pi\xi m
\end{equation}
in a manner similar to Eq.~(\ref{shift}).  If the fermion mass is
negative and large enough, the cosine term can dominate the behavior
around small $\eta^\prime$, making the perturbative vacuum unstable.
The bosonization process relates $\overline\psi\gamma_5\psi$ with
$\sin(2\sqrt\pi\eta^\prime)$; thus, when $\eta^\prime$ gains an
expectation value, so does the the pseudo-scalar density.  Since the
scalar field represents the electric field, this symmetry breaking
represents the spontaneous generation of a background field.  As
discussed by Coleman \cite{Coleman:1975pw,Coleman:1976uz}, this
corresponds to a non-trivial topological term in the action, usually
referred to as $\Theta$.

Another way to understand the one flavor behavior is to consider
several flavors and give all but one large masses.  For example
consider the three flavor case, and give two quarks a larger mass than
the third.  Model the light pseudoscalar sector of this theory with an
effective field $\Sigma$ taken from the group $SU(3)$.  With two
quarks of mass $M$ and one of mass $m$, consider the potential
\begin{equation}
V(\Sigma)\propto -{\rm ReTr}({\cal M}\Sigma)
\end{equation}
with mass matrix
\begin{equation}
{\cal M}=
\pmatrix{
m & 0 & 0\cr
0 & M & 0\cr
0 & 0 & M\cr
}.
\end{equation}
It is convenient to break this into two terms
\begin{equation}
V(\Sigma) \propto -{M+m\over 2}\ {\rm ReTr} (\Sigma)
+{M-m\over 2}\ {\rm ReTr}(\Sigma h)
\label{competing}
\end{equation}
where
\begin{equation}
h=\pmatrix{
1 & 0 & 0\cr
0 & -1 & 0\cr
0 & 0 & -1\cr
}.
\end{equation}
When $M+m$ is positive The minimum of the first term in
Eq.~(\ref{competing}) occurs at the identity element.  For the second
term, note that the factor $h$ of $\Sigma$ is taken as an $SU(3)$
group element.  The extrema of this term occur when the product
$\Sigma h$ is in the group center.  For the case $M-m$ positive, there
is a degenerate pair of minima occurring at
\begin{equation}
\Sigma=e^{\pm 2\pi i/3}h.
\end{equation}
We see that Eq.~(\ref{competing}) has two competing terms, one having
a unique minimum at $\Sigma=I$ and the other having two degenerate
ground states at the above complex values.  For the degenerate case
with $M=m$, only the first term is present and the vacuum is unique.
However when $m=-M$ only the second term is present with its
corresponding pair of degenerate vacua.  Somewhere between these
points must lie a critical value $m_c$ where the situation shifts
between a unique and a doubly degenerate vacuum.

To determine the critical mass, consider matrices of form
\begin{equation}
\Sigma=e^{i\phi\Gamma}
\end{equation}
where 
\begin{equation}
\Gamma=
\pmatrix{ -2 & 0 & 0\cr 0 & 1 & 0\cr 0 & 0 & 1\cr }.
\end{equation}
For these the potential is
\begin{equation}
V(\phi) \propto -m\cos(2\phi)-2M\cos(\phi).
\end{equation}
The extremum at $\phi=0$ continuously changes from a minimum to a
maximum at $m_c=-M/2$, the desired critical point.  As discussed at
the beginning of this section, it occurs at a negative value of $m$.
The only dimensional scale present is $M$, to which the result must be
proportional.  This analysis immediately generalizes to larger flavor
groups; for $N_f$ flavors divided into a set of $N_f-1$ of mass $M$
and one of mass $m$ we have $m_c={-M\over N_f-1}$.  As $M$ becomes
much larger than the scale of QCD, $\Lambda_{qcd}$, it is expected
that the latter will replace $M$ in setting the scale for the critical
mass.

This discussion suggests that a similar phenomenon should occur on the
lattice with one flavor of Wilson fermion.  Here the bare mass is
controlled by the hopping parameter.  As the hopping parameter
increases, the fermion mass decreases.  In the plane of the gauge
coupling and hopping parameter, a critical line should mark where the
above parity breaking begins.  In the lattice context the possibility
of such a phase was mentioned briefly by Smit \cite{Smit:1980nf} and
later extensively discussed by Aoki \cite{Aoki:1987us} and Aoki and
Gocksch \cite{Aoki:1992nb}.  The latter papers also made some rather
dramatic predictions for the breaking of both parity and flavor
symmetries when more quark species are present.

Ref.~\cite{Vafa:1983tf} has argued that certain discrete symmetries
such as parity cannot be spontaneously broken in vector-like theories
such as QCD.  This argument, however, assumes that parameters are
taken such that the fermion determinant is positive.  This is not true
at negative quark mass, i.e. where these effective Lagrangian arguments
suggest these strange phases exist.  In more conventional language,
the spontaneous symmetry breaking considered here occurs when the
topological angle $\Theta$ takes the value $\pi$.

\section{Dirac eigenvalues}
\label{eigenvalues}
Amongst the lattice gauge community it has recently become quite
popular to study the distributions of eigenvalues of the Dirac
operator in the presence of the background gauge fields generated in
simulations.  There are a variety of motivations for this.  First, in
a classic work, Banks and Casher \cite{Banks:1979yr} related the
density of small Dirac eigenvalues to spontaneous chiral symmetry
breaking.  Second, lattice discretizations of the Dirac operator based
the Ginsparg-Wilson relation \cite{Ginsparg:1981bj} have the
corresponding eigenvalues on circles in the complex plane.  The
validity of various approximations to such an operator can be
qualitatively assessed by looking at the eigenvalues.  Third, using
the overlap method \cite{Neuberger:1997fp} to construct a Dirac
operator with good chiral symmetry has difficulties if the starting
Wilson fermion operator has small eigenvalues.  This can influence the
selection of simulation parameters, such as the gauge action
\cite{Aoki:2002vt}.  Finally, since low eigenvalues impede conjugate
gradient methods, separating out these eigenvalues explicitly can
potentially be useful in developing dynamical simulation algorithms
\cite{Duncan:1998gq}.  This section is similar in philosophy to
Ref.~\cite{Leutwyler:1992yt}, wherein effective Lagrangian arguments
similar to those in the previous section were used to make predictions
for the eigenvalue structure for various numbers of flavors.

Despite this interest in the eigenvalue distributions, dangers lurk
for interpreting the observations.  Physical results come from the
full path integral over both the bosonic and fermionic fields.  Doing
these integrals one at a time is fine, but trying to interpret the
intermediate results is inherently dangerous.  While the Dirac
eigenvalues depend on the given gauge field, it is important to
remember that in a dynamical simulation the gauge field distribution
itself depends on the eigenvalues.  This circular behavior gives a
highly non-linear system, and such systems are notoriously hard to
interpret.

To establish the context of the discussion.  Consider a generic path
integral for a gauge theory
\begin{equation}
Z=\int (dA)(d\psi)(d\overline\psi)\ e^{-S_G(A)+\overline\psi D(A) \psi}.
\end{equation}
Here $A$ and $\psi$ represent the gauge and quark fields,
respectively, $S_G(A)$ is the pure gauge part of the action, and
$D(A)$ represents the Dirac operator in use for the quarks.  As the
action is quadratic in the fermion fields, a formal integration gives
\begin{equation}
Z=\int (dA)\ |D(A)|\ e^{-S_G(A)}.
\label{path}
\end{equation}
Working on a finite lattice, $D(A)$ is a finite dimensional matrix,
and for a given gauge field we can formally consider its eigenvectors
and eigenvalues
\begin{equation}
D(A)\psi_i=\lambda_i\psi_i.
\end{equation}
The determinant appearing in Eq.~(\ref{path}) is the product of these
eigenvalues; so, the path integral takes the form
\begin{equation}
Z=\int (dA)\ e^{-S_G(A)}\ \prod_i \lambda_i.
\end{equation}
Averaging over gauge fields defines the eigenvalue density
\begin{equation}
\rho(x+iy)={1\over {N}Z}\int (dA)\ |D(A)|\ e^{-S_G(A)}
\sum_{i}\delta(x-{\rm Re}\lambda_i(A))\delta(y-{\rm Im}\lambda_i(A)).
\end{equation}
Here $N$ is the dimension of the Dirac operator, including volume,
gauge, spin, and flavor indices.  

In situations where the fermion determinant is not positive, $\rho$
can be negative or complex; nevertheless, we still refer to it as a
density.  In addition, assume that $\rho$ is real; situations where
this is not true, such as with a finite chemical
potential,\cite{Osborn:2005ss} are fascinating but beyond the scope of
this discussion.

\begin{figure*}
\centering
\includegraphics[width=3in]{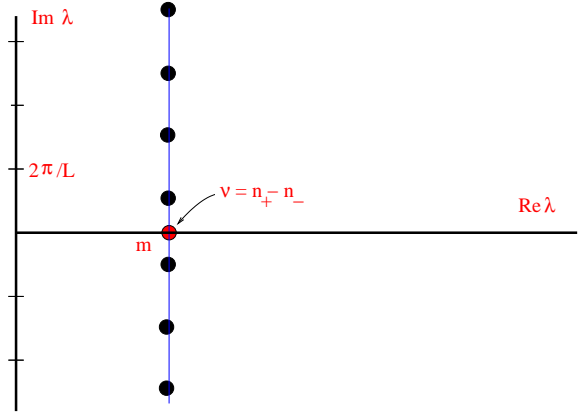}
\caption{ In the naive continuum picture, all eigenvalues of the Dirac
operator lie along a line parallel to the imaginary axis.  In a finite
volume these eigenvalues become discrete.  The real eigenvalues divide
into distinct chiralities and define a topological invariant.
\label{continuum}
}
\end{figure*} 

At zero chemical potential, all actions used in practice satisfy
``$\gamma_5$ Hermiticity''
\begin{equation}
\gamma_5 D \gamma_5=D^\dagger.
\label{hermiticity}
\end{equation}
With this condition, all non-real eigenvalues occur in complex
conjugate pairs, implying for the density
\begin{equation}
\rho(z)=\rho(z^*).
\end{equation}
This property will be shared by all the operators considered in the
following discussion.  

The quest is to find general statements relating the behavior of the
eigenvalue density to physical properties of the theory.  Repeating
the earlier warning, $\rho$ depends on the distribution of gauge
fields $A$ which in turn is weighted by $\rho$ which depends on the
distribution of $A$ \ldots.

\subsection{The continuum}

Of course the continuum theory is only really defined as the limit of
the lattice theory.  Nevertheless, it is perhaps useful to recall the
standard picture, where the Dirac operator
$$
D=\gamma_\mu (\partial_\mu+igA_\mu)+m
$$
is the sum of an anti-Hermitian piece and the quark mass $m$.  All
eigenvalues have the same real part $m$
$$
\rho(x+iy)=\delta(x-m) \tilde\rho(y).
$$ The eigenvalues lie along a line parallel to the imaginary axis,
while the Hermiticity condition of Eq.~(\ref{hermiticity}) implies
they are either real or occur in complex conjugate pairs.  Restricted
to the subspace of real eigenvalues, $\gamma_5$ commutes with $D$ and
thus these eigenvectors can be separated by chirality.  The difference
between the number of positive and negative eigenvalues of $\gamma_5$
in this subspace defines an index related to the topological structure
of the gauge fields \cite{index}. The basic structure is sketched in
Fig.~(\ref{continuum}).

It is useful to separately discuss the consequences of real versus
complex eigenvalues.  The former form a continuous distribution
whereas the latter are all at the same point.  This section
concentrates on the density of complex eigenvalues, the next section
turns to the consequences of the exactly real eigenvalues.

The Banks and Casher argument relates a non-vanishing $\tilde\rho(0)$
to the chiral condensate occurring when the mass goes to zero.  We
will say more on this later in the lattice context.  Note that the
naive picture suggests a symmetry between positive and negative mass.
Due to anomalies, this is spurious.  Indeed, with any number of
flavors, flipping the sign of a single quark mass gives an
inequivalent theory.

\begin{figure*}
\centering
\includegraphics[width=2.5in]{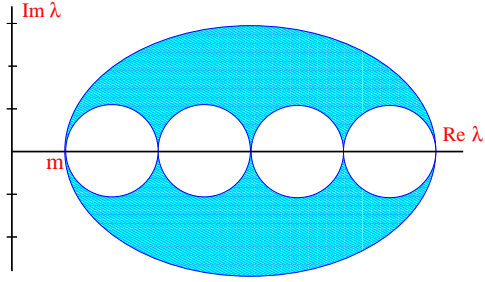}
\caption{ Free Wilson fermions display an eigenvalue spectrum with a
momentum dependent real part.  This removes doublers by giving them a
large effective mass.}
\label{fig2}

\end{figure*} 

\subsection{Wilson fermions}

The lattice reveals the true intricacy of the situation arising from
chiral anomalies.  Without ultraviolet infinities, all naive
symmetries of the lattice action are true symmetries.  Naive fermions
cannot have anomalies, which are cancelled by extra states referred to
as doublers.  Wilson fermions \cite{Wilson:1975id} avoid this issue by
giving a large real part to those eigenvalues corresponding to the
doublers.  In particular, by modifying the hopping term of naive
fermions, Wilson allowed the fermion mass to depend on momentum
\begin{equation}
m\rightarrow m+{1\over a} \sum_\mu (1-\cos(p_\mu a))
\end{equation}
thus giving the
doublers a mass of order $1/a$.  In momentum space, the free
Wilson-Dirac operator takes the form
\begin{equation}
D_w= m+{1\over a}\sum_\mu (i \sin(p_\mu a)\gamma_\mu+ 1-\cos(p_\mu a)).
\end{equation}
The corresponding eigenvalue structure displays a simple pattern as
shown in Fig.~(\ref{fig2}).

As the gauge fields are turned on, the eigenvalues shift around and
blur this pattern.  An additional complication is that the operator
$D$ is no longer normal, i.e. $[D,D^\dagger]\ne 0$ and the
eigenvectors need not be orthogonal.  The complex eigenvalues are
still paired, although, as the gauge fields vary, complex pairs of
eigenvalues can collide and separate along the real axis.  In general,
the real eigenvalues will form a continuous distribution.

As in the continuum, an index can be defined from the spectrum of the
Wilson-Dirac operator.  Again, $\gamma_5$ Hermiticity allows real
eigenvalues to be sorted by chirality.  To remove the contribution of
the doubler eigenvalues, select a point inside the leftmost open
circle of Fig.~(\ref{fig2}).  Then define the index of the gauge field
to be the net chirality of all real eigenvalues below that point.  For
smooth gauge fields this agrees with the topological winding number
obtained from their interpolation to the continuum.  It also
corresponds to the winding number discussed below for the overlap
operator.

\subsection{The overlap}

\begin{figure*}
\centering
\includegraphics[width=5in]{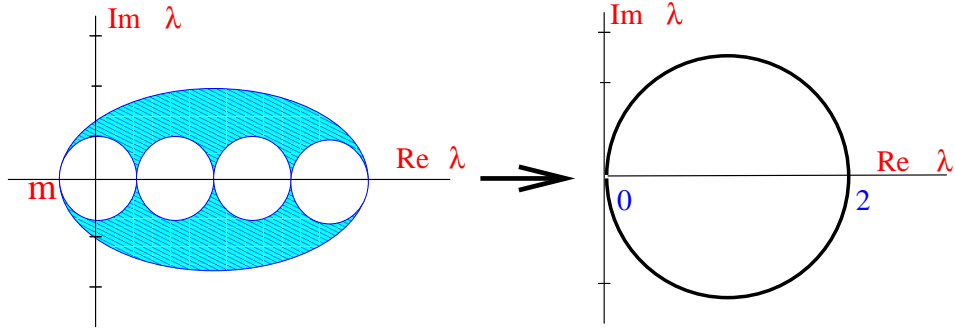}
\caption{The overlap operator is constructed by projecting the Wilson
  Dirac operator onto a circle.}
\label{fig3}
\end{figure*}

Wilson fermions have a rather complicated behavior under chiral
transformations.  The overlap formalism\cite{Neuberger:1997fp}
simplifies this by first projecting the Wilson matrix $D_W$ onto a
unitary operator
\begin{equation}
V=(D_W D_W^\dagger)^{-1/2} D_W.
\end{equation}
This is to be understood in terms of going to a basis that
diagonalizes $D_W D_W^\dagger$, doing the inversion, and then
returning to the initial basis.  In terms of this unitary quantity,
the overlap Dirac operator is
\begin{equation}
D=1+V.
\end{equation}
The projection process is sketched in Fig.~(\ref{fig3}).  The mass
used in the starting Wilson operator is taken to a negative value so
selected that the low momentum states are projected to low
eigenvalues, while the doubler states are driven towards $\lambda\sim
2$.

The overlap operator has several nice properties.  First, it satisfies
the Ginsparg-Wilson relation,\cite{Ginsparg:1981bj} 
\begin{equation}
\gamma_5 D + D\gamma_5=D\gamma_5 D
\end{equation}
which is now most succinctly written as the unitarity of $V$ coupled
with its $\gamma_5$ Hermiticity
\begin{equation}
\gamma_5 V\gamma_5 V=1.
\end{equation}
As it is constructed from a unitary operator, normality of $D$ is
guaranteed.  But, most important, it exhibits a lattice version of an
exact chiral symmetry \cite{Luscher:1998pq}.  The fermionic action
$\overline\psi D\psi$ is invariant under the transformation
\begin{equation}
\matrix{
&\psi\rightarrow e^{i\theta\gamma_5}\psi\cr
&\overline\psi\rightarrow \overline\psi e^{i\theta\hat\gamma_5}\cr
}
\label{symmetry}
\end{equation}
where
\begin{equation}
\hat\gamma_5 =-V\gamma_5.
\end{equation}
As with $\gamma_5$, this quantity is Hermitean and its square is unity.
Thus its eigenvalues are all plus or minus unity.  The trace
defines an index
\begin{equation}
\nu={1\over 2}{\rm Tr}\hat\gamma_5
\end{equation}
which plays exactly the role of the index in the continuum.
The factor of one half in this equation is due to the fact that the
total number of real eigenvalues is even with each zero
eigenvalue of $D$ having a partner at $D=2$, and both contribute to
$\hat\gamma_5$. 

Of course for the one flavor theory anomalies remove all traces of
chiral symmetry; so, the use of the overlap operator in this case
seems less motivated.  Nevertheless, the formalism has the nice
properties of having the Dirac operator be normal and of having exact
zero modes.  This allows an analysis of how, despite this apparent
extra symmetry, the eigenvalue structure still permits the predicted
smooth behavior for this theory around zero mass.

It is important to note that the overlap operator is not unique.  Its
precise form depends on the particular initial operator chosen to
project onto the unitary form.  Using the Wilson-Dirac operator for
this purpose, the result still depends on the input mass used.  From
its historical origins in the domain wall formalism, this quantity is
sometimes called the ``domain wall height.''

Because the overlap is not unique, an ambiguity can remain in
determining the winding number of a given gauge configuration.  Issues
arise when $D_W D_W^\dagger$ is not invertible, and for a given gauge
field this can occur at specific values of the projection point.
This problem can be avoided for ``smooth'' gauge fields.  Indeed, an
``admissibility condition,''~\cite{Luscher:1981zq,Hernandez:1998et}
requiring all plaquette values to remain sufficiently close to the
identity, removes the ambiguity.  Unfortunately this condition is
incompatible with reflection positivity \cite{Creutz:2004ir}.  Because
of these issues, it is not known if the topological susceptibility is
in fact a well defined physical observable.  On the other hand, as it
is not clear how to measure the susceptibility in a scattering
experiment, there seems to be little reason to care if it is an
observable or not.

To control issues related to exact zero modes, introduce a small mass
and take the volume to infinity first and then the mass to zero.
Toward this end, consider
\begin{equation}
\langle\overline\psi\psi\rangle=\langle {\rm Tr}\ D^{-1}\rangle
=\left\langle\sum_i {1\over\lambda_i+m}\right\rangle.
\end{equation}
The signal for chiral symmetry breaking is a jump in this quantity as
the mass passes through zero.

\begin{figure*}
\centering
\includegraphics[width=3in]{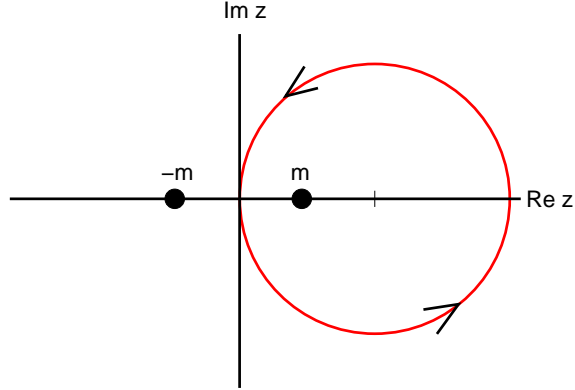}
\caption{As the mass changes sign a pole moves between inside and outside
the overlap circle.  This generates a jump in the condensate.}
\label{circles2}
\end{figure*}

As the volume goes to infinity, replace the above sum with a contour
integral around the overlap circle using $\lambda=1+e^{i\theta}$.  Up
to the trivial volume factor, one should evaluate
\begin{equation}
i\int_0^{2\pi} d\theta {\rho(\theta)\over 1+e^{i\theta}+m}.
\end{equation}
As the mass passes through zero, the pole at $\lambda=-m$ passes
between lying outside and inside the circle, as sketched in
Fig.~(\ref{circles2}).  As it passes through the circle, the residue
of the pole is $\rho(0) = \lim_{\theta\rightarrow 0}\rho(\theta)$.
Thus the integral jumps by $2\pi\rho(0)$.  This is the overlap version
of the Banks-Casher relation \cite{Banks:1979yr}; a non-trivial jump
in the condensate is correlated with a non-vanishing $\rho(0)$.

Taking the volume to infinity before taking the mass to zero is
important here.  On any finite volume the partition function is a
finite and convergent integral which is analytic in the mass and there
can be no phase transition.

Note that for multiple flavors the exact zero modes related to
topology are suppressed by the mass and do not contribute to this
jump.  For one flavor, however, the zero modes do give rise to a
non-vanishing but smooth contribution to the condensate, as discussed
in the next section.\cite{Damgaard:1999ij}

There is an interesting contrast between the one flavor theory and the
picture when there are multiple degenerate quarks.  For example, with
two flavors of light quarks one expects spontaneous chiral symmetry
breaking.  This is the explanation for the light mass of the pion,
which is an approximate Goldstone boson.  In the above picture, the
two flavor theory should have a non-vanishing $\rho(0)$.

Now return to the one flavor theory.  In this case there should be no
chiral symmetry.  The famous $U(1)$ anomaly breaks the naive symmetry.
No massless physical particles are expected when the quark mass
vanishes.  Furthermore, the earlier simple chiral Lagrangian arguments
\cite{DiVecchia:1980ve,Creutz:2003xu} indicate that no singularities
are expected when a single quark mass passes through zero.  Combined
with the above discussion, this leads to the conclusion that for the
one flavor theory $\rho(0)$ must vanish.

This leads us to an amusing paradox.  Consider the original path
integral after the fermions are integrated out.  Changing the number
of flavors $N_f$ manifests itself in the power of the determinant
\begin{equation}
\int dA\  |D|^{N_f}\  e^{-S_g(A)}.
\end{equation}
Naively this suggests that as you increase the number of flavors, the
density of low eigenvalues should decrease.  But we have just argued
that with two flavors $\rho(0)\ne 0$ but with one flavor $\rho(0)= 0$.
How can it be that increasing the number of flavors actually increases
the density of small eigenvalues?

This is a clear example of how the non-linear nature of the problem
can produce non-intuitive results.  The eigenvalue density depends on
the gauge field distribution, but the gauge field distribution depends
on the eigenvalue density.  It is not just the low eigenvalues that
are relevant to the issue.  Fermionic fields tend to smooth out gauge
fields, and this process involves all scales.  Smoother gauge fields
in turn can give more low eigenvalues.  Thus high eigenvalues
influence the low ones, and this effect evidently can overcome the
naive suppression from more powers of the determinant.

\section{Zero modes}
\label{zeromodes}
Through the index theorem, the topological structure of the gauge
field manifests itself in zero modes of the massless Dirac operator.
These are closely tied to the chiral anomaly and the behavior of the
quark condensate for small quark masses.  In this section I further
explore this connection in the overlap formalism, concentrating on
these zero modes.

As before, integrating out the fermionic fields from the path integral
gives a determinant of the Dirac operator, $D$.  For any given
configuration of gauge fields this determinant is the product of the
eigenvalues of this matrix.  To control infrared issues, insert a
small mass and write the path integral
\begin{equation}
Z=\int dA\ 
e^{-S_g}\ 
\prod_i (\lambda_i+m). 
\end{equation}
Here the $\lambda_i$ are the eigenvalues of the kinetic part of the
fermion determinant.  If we take the mass to zero, any configurations
which contain a zero eigenmode will have zero weight in the path
integral.  This suggests that for the massless theory we can ignore
any instanton effects since those configurations don't contribute to
the path integral.

What is wrong with this argument?  The issue is not whether the zero
modes contribute to the path integral, but whether they can contribute
to physical correlation functions.  To see how this goes, add some sources
to the path integral
\begin{equation}
Z(\eta,\overline\eta)=\int dA\ d\psi\ d\overline\psi\  
e^{-S_g+\overline\psi (D+m) \psi +\overline\psi \eta+ \overline\eta\psi}.
\end{equation}
Differentiation (in the Grassmann sense) with respect to $\eta$ and
$\overline \eta$ gives any desired fermionic correlation function.
Now integrate out the fermions
\begin{equation}
Z=\int dA\ 
e^{-S_g-\overline\eta(D+m)^{-1}\eta}\ 
\prod_i (\lambda_i+m).
\end{equation}
If we consider a source that overlaps with an eigenvector of $D$
corresponding to one of the zero modes, i.e.
\begin{equation}
(\psi_0,\eta)\ne 0,
\end{equation}
the source contribution introduces a $1/m$ factor.  This cancels the
$m$ from the determinant, leaving a finite contribution as $m$ goes to
zero \cite{Damgaard:1999ij}.

With multiple flavors, the determinant will have a mass factor from
each.  When several masses are taken to zero together, one will need a
similar factor from the sources for each.  This product of source
terms is the famous ```t Hooft vertex.''
\cite{'tHooft:1976up,'tHooft:fv} While it is correct that instantons
do drop out of $Z$, they survive in correlation functions.

While these issues are well understood theoretically, they can raise
potential difficulties for numerical simulations.  The usual procedure
generates gauge configurations weighted as in the partition function.
For a small quark mass, topologically non-trivial configurations will
be suppressed.  But in these configurations, large correlations can
appear due to instanton effects.  This combination of small weights
with large correlations can give rise to large statistical errors,
thus complicating small mass extrapolations.  The problem will be
particularly severe for quantities dominated by anomaly effects, such
as the $\eta^\prime$ mass.  A possible strategy to alleviate this
effect is to generate configurations with a modified weight, perhaps
along the lines of multi-canonical algorithms.\cite{Berg:1992qu}

In our case of the one flavor theory, the 't Hooft vertex is a
quadratic form in the fermion sources.  This will give a finite
contribution to the condensate $\langle\overline\psi\psi\rangle$ that
is continuous in the mass as the mass passes through zero.  Note that
unlike the jump generated from complex eigenvalues discussed in the
previous section, this contribution remains even at finite volume.  As
the volume goes to infinity it is still only the one instanton sector
that contributes since all instantons far from the source
$\overline\psi\psi$ get suppressed by the mass factor.

Indeed, the 't Hooft vertex represents a non-perturbative additive
shift to the quark mass \cite{mcarthurgeorgi}.  As discussed in the
next section, the size of this shift generally depends on scale and
regulator details.  Even with the Ginsparg-Wilson condition, the
lattice Dirac operator is not unique, and there is no proof that two
different forms have to give the same continuum limit for vanishing
quark mass.  Because of this, the concept of a single massless quark
is not physical \cite{Creutz:2004fi}, invalidating one popular
proposed solution to the strong CP problem.  This ambiguity has been
noted for heavy quarks in a more perturbative context
\cite{Bigi:1994em} and is often referred to as the ``renormalon''
problem.

\section{The renormalization group}
\label{rgroup}

In previous sections we discussed the quark mass as a simple parameter
without really defining it precisely.  Because of confinement, quarks
are not free particles and the usual definition of mass via particles
propagating over long distances does not apply.  Furthermore, as we
are dealing with a quantum field theory, all bare parameters are
divergent and need renormalization.  In this section we use
renormalization group methods to accomplish this, giving a precise
definition to a quark mass in the context of a given scheme.  This
will expose certain ambiguities in the mass definition.  Most of this
section is an expansion on the ideas presented in
Ref.~\cite{Creutz:2004fi}.

The renormalization process tunes all relevant bare parameters as a
function of the cutoff while fixing a set of renormalized quantities.
As we need to renormalize both the bare coupling and quark mass, we
need to fix two physical observables.  For this purpose, choose the
lightest boson and the lightest baryon masses.  As both are expected
to be stable, this precludes any ambiguity from particle widths.  As
before, we denote the lightest boson as the ${\eta^\prime}$.  In the
one flavor theory the lightest baryon $p$ actually has spin 3/2 due to
Pauli statistics, but for simplicity we still refer to it as the
``proton.''  Because of confinement, the values of their masses are
inherently non-perturbative quantities.

With the cutoff in place, the physical masses are functions of
$(g,m,a)$, the bare charge, the bare coupling, and the cutoff.  For
simplicity in this section, consider only real $m$ and ignore the CP
violating mass term $m_5$.  Holding the masses constant, the
renormalization process determines how $g$ and $m$ flow as the cutoff
is removed.  Because of asymptotic freedom, this flow eventually
enters the perturbative regime and we have the famous renormalization
group equations \cite{Gell-Mann:fq}
\begin{eqnarray}
a{dg\over da}&\equiv&\beta(g)=\beta_0 g^3+\beta_1 g^5 +\ldots \\
a{dm\over da}&\equiv&m\gamma(g)=m(\gamma_0 g^2+\gamma_1 g^4 +\ldots)
+{\rm non\hbox{-}perturbative}.
\end{eqnarray}
The ``non-perturbative'' term should vanish faster than any power of
the coupling.  We include it explicitly in the mass flow because it
will play a crucial role in the latter discussion.  The values for the
first few coefficients $\beta_0$, $\beta_1$, and $\gamma_0$ are known
\cite{coef} and independent of regularization scheme.

\begin{figure*}
\centering
\includegraphics[width=2.5in]{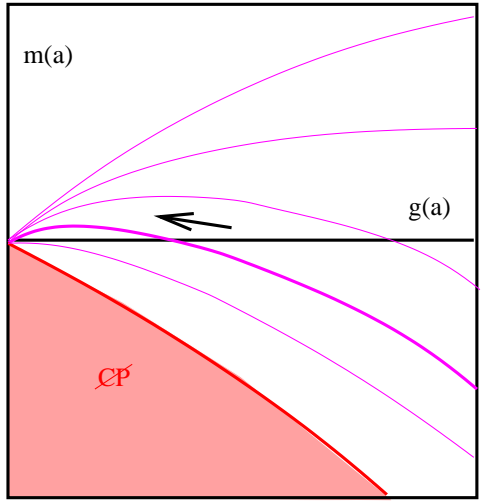}
\caption{As the cutoff is removed, the bare coupling and mass flow
 towards zero.  Different flow lines correspond to different
 renormalized quark masses.  With sufficiently negative mass one
 enters the regime of spontaneous CP violation.  Anomaly effects
 preclude the line of vanishing bare quark mass from being
 renormalization group invariant.  }
\label{rgflow}
\end{figure*}

The solution to these equations is well known
\begin{eqnarray}
a&=&{1\over \Lambda_{qcd}} e^{-1/2\beta_0 g^2} g^{-\beta_1/\beta_0^2}
(1+O(g^2))\label{flow1}\\
m&=&Mg^{\gamma_0/\beta_0}
(1+O(g^2)).
\label{flow2}
\end{eqnarray}
In particular this shows how the bare coupling and bare
mass are driven to zero as the cutoff is removed
\begin{eqnarray}
g&\sim& {1\over\log(1/a\Lambda_{qcd} )}\\
m&\sim& {1\over(\log(1/a\Lambda_{qcd} ))^{\gamma_0/\beta_0}}.
\end{eqnarray}
This flow is sketched schematically in Fig.~(\ref{rgflow}).

The quantities $\Lambda_{qcd}$ and $M$ are integration constants for
the renormalization group equations.  We refer to $\Lambda_{qcd}$ as
the overall strong interaction scale and $M$ as the renormalized quark
mass.  Their values depend on the explicit renormalization scheme as
well as the physical values being held fixed in the renormalization
process, i.e. the proton and $\eta^\prime$ masses.  This connection is
highly non-perturbative.  Indeed the particle masses being fixed are
long distance properties, and thus are tied to the renormalization
group flow far out of the perturbative regime.

Turning things around, we can consider the physical particle masses as
functions of these integration constants.  Simple dimensional analysis
tells us that the dependence of physical masses must take the form
\begin{eqnarray}
m_p&=&\Lambda_{qcd} H_p(M/\Lambda_{qcd})\\
m_{\eta^\prime}&=&\Lambda_{qcd} H_{\eta^\prime}(M/\Lambda_{qcd})
\end{eqnarray}
where the $H_i(x)$ are dimensionless functions whose detailed form is
highly non-perturbative.

For the case of multiple degenerate quark flavors we expect the square
of the lightest boson mass to vanish linearly as the renormalized
quark mass goes to zero.  This means we anticipate a square root
singularity in the corresponding $H(x)$ at $x=0$.  Indeed, requiring
the singularity to occur at the origin removes any additive
non-perturbative ambiguity in defining the renormalized mass.

For the one flavor theory, things are more subtle.  As discussed
above, we expect physics to behave smoothly as the quark mass passes
through zero.  That is, we do not expect $H_i(x)$ to display any
singularity at $x=0$.  Non-perturbative dynamics generates an
additional contribution to the mass of the pseudo-scalar meson; thus,
the $M=0$ flow generically corresponds to a positive value of
$m_{\eta^\prime}$.  While a $m_{\eta^\prime}=0$ flow line can exist,
it represents the boundary of the CP violating phase discussed earlier
and has little to do with massless quarks.

Now we come to the question of scheme dependence.  Given some
different renormalization prescription, i.e. a modified lattice
action, the precise flows will change.  Although the behavior dictated
in Eqs.~(\ref{flow1},\ref{flow2}) must be preserved, the integration
constants $(\Lambda_{qcd},M)$ and the functions $H_i(x)$ will in
general be modified.  Marking the new quantities with tilde's,
matching the schemes to give the same physics requires
\begin{equation}
m_i=\Lambda_{qcd} H_i(M/\Lambda_{qcd})=\tilde \Lambda_{qcd} 
\tilde H_i(\tilde M/\tilde\Lambda_{qcd}).
\end{equation}
Upon the removal of the cutoff, two different valid schemes should
give the same result for the physical masses.  An important
distinction for the one flavor theory, without chiral symmetry to
protect things, is the absence of any reason for the vanishing of $M$
to require the vanishing of $\tilde M$.

On changing schemes, we introduce new definitions for the coupling and
mass.  To match onto the perturbative limit, it is reasonable to
restrict these definitions to agree at leading order.  Thus we should
require
\begin{eqnarray}
\tilde g&=& g+O(g^3)\label{pertmatch1}\\
\tilde m&=& m(1+O(g^2))+\hbox {non-perturbative.}
\label{pertmatch2}
\end{eqnarray}
Here the ``non-perturbative'' terms should vanish faster than any
power of the coupling, but are not in general proportional to $m$.  In
particular, a non-perturbative additive shift in the quark mass
follows qualitatively from the analysis of zero modes in the previous
section.

The requirements for the perturbative limit apply at fixed cutoff.
Indeed, the interplay of the $a\rightarrow 0$ and the $g\rightarrow 0$
limits is rather intricate.  As $g\rightarrow 0$ at fixed $a$ the
quarks decouple and we have a theory of free quarks and gluons.  The
limit $a\rightarrow 0$ at fixed $g$ brings on the standard divergences
of relativistic field theory.  The proper continuum limit follows the
renormalization group trajectory with both $a$ and $g$ going together
in the appropriate way and gives a theory where important
non-perturbative effects such as confinement are relevant.

Assuming only the matching conditions in
Eq.~(\ref{pertmatch1},\ref{pertmatch2}) leaves the freedom to do some
amusing things.  As a particularly contrived example, consider
\begin{eqnarray}
\tilde g &=& g\\
\tilde m&=&m-M g^{\gamma_0/\beta_0}\times
{ e^{-1/2\beta_0 g^2} g^{-\beta_1/\beta_0^2}\over \Lambda_{qcd} a}.
\end{eqnarray}
The last factor vanishes than any power of $g$, but is crafted to go
to unity along the renormalization group trajectory.  Note that a
power of the scale factor as inserted here is necessary for
non-perturbative phenomena to be relevant to the continuum limit
\cite{'tHooft:1976up}.  With this form, one can immediately relate the
old and new renormalized masses
\begin{equation}
\tilde M\equiv
\lim_{a\rightarrow 0}\ \tilde m \tilde g^{-\gamma_0/\beta_0} = 
M-M=0.
\end{equation}
Thus for any $M$, another scheme always exists where the renormalized
quark mass vanishes.  The possibility of such a transformation
demonstrates that masslessness is not a physical concept for the one
flavor theory, or more generally for a non-degenerate quark in a
multi-flavor theory.

I close this section with some remarks on trying to define the quark
mass is through the operator product expansion.  Expanding the product
of two electromagnetic currents at small separations will bring in a
variety of quark operators.  Among them is the simple scalar
combination $\overline\psi\psi$.  The short distance behavior involves
a triangle diagram which by $\gamma_5$ symmetry should vanish when the
quark mass does.  Thus one might define the zero mass theory where the
coefficient of the singular part of this term in the operator product
expansion has a zero.

This approach raises several issues.  Because the 't Hooft vertex in
the one flavor theory takes the same form $\overline\psi\psi$, it
clouds the definition of this as a renormalized operator.  The same
scheme dependent additive shift that plagues the quark mass can modify
where this zero occurs.  Furthermore, the additive shift in the quark
mass makes it unclear whether this definition of vanishing quark mass
has any connection with the quark masses in some particular effective
chiral Lagrangian.  It is even possible that the zero mass theory
defined this way may be in the CP violating phase, in which case it
certainly doesn't represent something physically interesting.  Indeed,
it is unclear what experiment if any could determine if the quark mass
defined via the operator product expansion vanishes.

\section{More colors and higher representations}
\label{colors}

A popular approximation considers QCD in the limit of a large number
of colors, replacing the $SU(3)$ gauge group with $SU(N_c)$.  In the
limit $N_c\rightarrow\infty$ planar diagrams dominate and internal
quark loops are suppressed \cite{'tHooft:1973jz}.  Unfortunately, most
of the effects being discussed in this paper are suppressed as well,
being higher order corrections in the $1/N_c$ expansion.
Nevertheless, an early indication of the basic one flavor behavior
came during a study of the large $N_c$ limit \cite{DiVecchia:1980ve}.
Thus it seems reasonable that for larger but finite $N_c$ and
retaining the fundamental representation of the gauge group for the
fermions, we should have a similar structure to that of
Fig.~(\ref{oneflavor}).

Recently a rather interesting variation of the large number of colors
expansion has been proposed
\cite{Corrigan:1979xf,Armoni:2003fb,Unsal:2006pj}.  Rather than taking
the quarks in the fundamental representation, they use the
antisymmetric tensor product of two fundamental representations.  For
the case $N_c=3$ these theories are equivalent, since it is a
convention whether quarks are in the 3 or the $\overline 3$
representation of $SU(3)$.  When $N_c$ increases, however, the product
representation is larger, of dimension $N_c(N_c-1)/2$ rather than
$N_c$, and enhances the effects of quark loops.  As the number of
colors goes to infinity the distinction between the antisymmetric and
symmetric tensor product becomes unimportant and the papers in
Ref.~\cite{Armoni:2003fb} have gone on to make inferences between the
these theories at $N_c=\infty$ and the bosonic sector of
supersymmetric Yang-Mills theory.  Thus they suggested using this
variation on the large color expansion to extract information about
one flavor QCD.

Working with fermions in a representation other than the fundamental
modifies the effect of topological structures.  In particular,
although there still is no continuous chiral symmetry in the one
flavor theory, certain discrete chiral symmetries can arise.  It is
then natural to ask if these discrete symmetries could be
spontaneously broken.  Ref.~\cite{Armoni:2003fb} suggests that they
are, and $\langle\overline\psi\psi\rangle$ represents an order
parameter for this breaking.  Evaluating this in the large $N_c$
limit, they propose that this might give some approximate information
on the theory with a smaller number of colors.  Of course for the
three color case of interest, there is no such extra symmetry to be
broken and it makes no sense to consider
$\langle\overline\psi\psi\rangle$ as an order parameter in the
traditional sense.  But these discrete symmetries are interesting in
their own right and it is perhaps instructive to ask if they are
indeed broken spontaneously for larger $N_c$.

The extra symmetries arise in cases where the zero modes of the Dirac
operator are automatically degenerate.  This is the case for the gauge
group $SU(N)$ for $N>3$ and fermions in the antisymmetric tensor product of two
fundamental representations.  Consider an instanton configuration, and
rotate it to appear in the $SU(2)$ subgroup involving only the first
two colors.  If we break the antisymmetric representation into
multiplets under this $SU(2)$ subgroup, it will consist of one singlet
from both indices being in the first two colors, $N-2$ doublets with
only one index from the first two and finally $(N-2)*(N-3)/2$
additional singlets involving only the higher values for the indices.
From the instanton we expect one zero mode for each of the doublets,
giving $N-2$ overall.  From the earlier discussion, this means the 't
Hooft vertex will involve the product of $N-2$ fermion bilinears.

Now consider the basic $U(1)$ chiral rotation of
Eq.~(\ref{chiralrot}).  The 't Hooft vertex still violates this
symmetry; however, if the angle $\theta$ is chosen to be a
multiple of ${2\pi\over N_c-2}$, this vertex remains invariant.  Thus
this theory has a hidden discrete $Z_{N_c-2}$ chiral symmetry.  The
conventional angle $\Theta$ conjugate to the gauge field topology
differs from the phase of the mass term by a factor of $N_c-2$.

This discussion assumes that all anomaly effects arise through
topological structures and the 't Hooft vertex.  On the lattice one
might also expect lattice artifacts to break the symmetry, much as
they do for conventional chiral symmetry in the Wilson action.  Even
for the overlap, there can be rough gauge configurations on which the
winding number is ill defined.  We assume that these issues involve
higher dimensional ``irrelevant'' operators and they disappear in the
continuum limit.

\begin{figure*}
\centering
\includegraphics[width=3in]{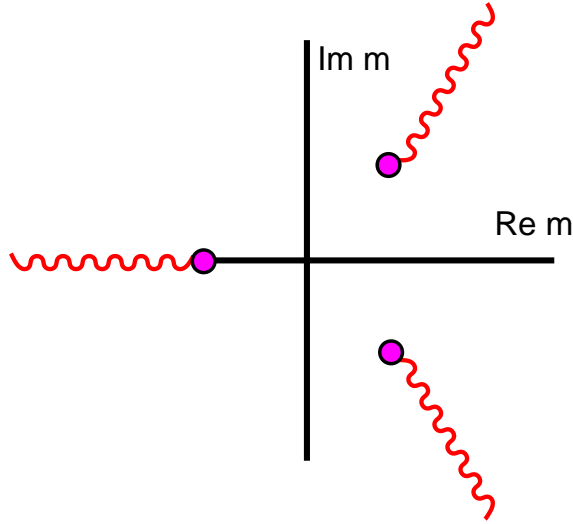}
\caption{ A possible phase diagram for one flavor quark-gluon dynamics
with gauge group $SU(5)$ and the fermions in the 10 representation.
Unlike the $SU(3)$ case, there are now three first order phase
transitions all pointing at the origin.  The transition along the
negative real axis represents a spontaneous breaking of CP symmetry.
As the number of colors increases, additional transition lines should
appear, with endpoints converging to the origin at $N_c=\infty$ where
there is a spontaneously broken $U(1)$ symmetry.  }
\label{sufive} 
\end{figure*}

Turning on a complex quark mass, the theory is invariant under
multiplication of this mass by an element of $Z_{N_c-2}$.  Thus the
phase diagram of Fig.~(\ref{oneflavor}) must be modified to
incorporate this symmetry.  To be specific, consider $SU(5)$ with the
fermions in the 10 dimensional representation.  (I skip over $SU(4)$
with fermions in the 6 dimensional representation to avoid the
complication of baryons being bosons made up of only two
quarks.)  By the above arguments the five color theory has a $Z_3$
discrete chiral symmetry.  One simple modification of the phase
diagram for complex mass that incorporates this symmetry is sketched
in Fig.~(\ref{sufive}).

\begin{figure*}
\centering
\includegraphics[width=2.5in]{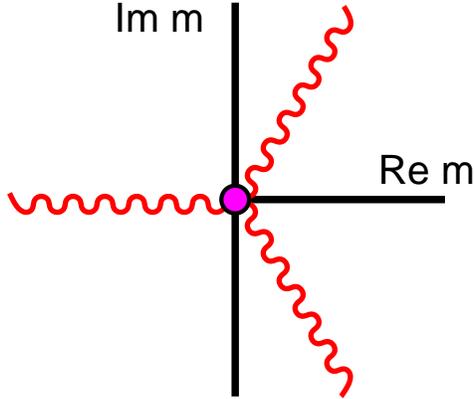}
\caption{ An alternative possible phase diagram for one flavor
quark-gluon dynamics with gauge group $SU(5)$ and the fermions in the
10 representation.  This represents the situation where the three
first order transitions meet at a triple point at the origin.  In this
case the discrete $Z_3$ chiral symmetry is spontaneously broken.  }
\label{sufivealt} 
\end{figure*}

As drawn, this figure assumes that this $Z_3$ symmetry is unbroken.
An alternative possibility is that it is spontaneously broken.  In
this case the three transition lines would extend to the origin.  The
order parameter for these transition lines is the expectation of the
$\eta^\prime$ field, which should undergo a finite jump as one passes
through them.  Given the highly suppressed contributions of instantons
in this theory, and in light of the smooth behavior of the three color
theory when the mass vanishes, this seems rather unmotivated at small
$N_c$, but the possible alternate phase diagram is sketched in
Fig.~(\ref{sufivealt}).  It appears possible that at some large but
finite $N_c$ there is a change in behavior from that exemplified by a
unique vacuum and transitions not reaching the origin, as in
Fig.~\ref{sufive}, to the case where there are $N_c-2$ degenerate
vacua at vanishing mass as in Fig.~\ref{sufivealt}.

Whether the discrete $Z_{N_c-2}$ symmetry is broken or not, the point
where the mass vanishes is now well defined, as long as one uses a
regulator which respects this discrete symmetry.  At this point
$\langle\overline\psi\psi\rangle$ either vanishes or shows first order
jumps to the other phases.  This contrasts with the usual QCD case
with gauge group $SU(3)$ where there is no residual discrete symmetry
and this ``condensate'' has a smooth and non-vanishing behavior.  As
the mechanism for generating this expectation is rather different in
the $N_c=3$ case from the spontaneous symmetry breaking for more
colors, it is unclear whether there should be any numerical connection
\cite{DeGrand:2006uy}.

As $N_c$ increases, there should be a corresponding growth in the
number of transition lines in the complex mass plane.  In terms of the
usual angle $\Theta$ appearing for topologically non-trivial gauge
configurations, these transitions are all equivalent and represent
$\Theta=\pi$.  The convergence of these lines towards the origin can
potentially give rise to the spontaneously broken $U(1)$ chiral
symmetry expected in the $N_c\rightarrow\infty$ limit.

Note that this situation contrasts sharply with the behavior of QCD
with several degenerate flavors.  There it is also true that the
topologically defined phase $\Theta$ differs by an integer factor from
the phase of the quark mass.  In the complex mass plane there are also
expected to be several transition lines converging on the origin
\cite{Creutz:2000bs,Creutz:1995wf}.  However in this case there are
massless Goldstone bosons when the mass exactly vanishes.

Considering other higher fermion representations, similar discrete
symmetries are expected, such as a $Z_5$ for color $SU(3)$ with
fermions in the $6$ representation.  For the adjoint case, each zero
mode is $2N_c$ degenerate and we have a discrete $Z_{2N_c}$ chiral
symmetry, although the meaning of confinement in this theory is
obscured since gluons can screen individual quarks.  Going still
further, one has to worry about whether one enters a conformal phase
and/or asymptotic freedom is lost.

\section{Conclusions}
\label{conclude}
One flavor QCD is a fascinating system.  Chiral symmetry, which is so
crucial to our conventional understanding of the strong interactions,
plays a rather strange role here.  Indeed, anomalies mean that the
naive classical chiral symmetry must disappear from the problem.  This
paper has discussed the qualitative behavior of the one flavor theory
as a function of the quark mass.  As summarized in
Fig.~(\ref{oneflavor}), a second order phase transition is expected at
non-zero negative mass.  At this point the $\eta^\prime$ mass
vanishes, while for still more negative mass this field acquires an
expectation value, marking a CP violating phase.

This picture enables partial answers to many of the questions raised
in the introduction.  Indeed, chiral symmetry is in some sense
irrelevant to the one flavor theory.  Physics varies smoothly and
continuously for small masses and the location of the $m=0$ point is
not well defined.  The quark condensate
$\langle\overline\psi\psi\rangle$ is automatically non-zero and ceases
to be a natural order parameter for any broken symmetry.  However,
with more colors and quarks in higher representations than the
fundamental, discrete chiral symmetries can emerge for which the
condensate may be an order parameter.  However, it is an open question
when these symmetries are expected to be broken for a finite number of
colors.

Many of these details are in principle amenable to study in numerical
simulations.  Such simulations are made more difficult by the small
mass region, and involve sign problems when the quark mass is
negative.  The latter will become particularly severe near the phase
of spontaneously broken CP.  Nevertheless, the absence of massless
Goldstone bosons should alleviate these problems relative to theories
with more flavors.

\section*{Acknowledgments}
This manuscript has been authored under contract number
DE-AC02-98CH10886 with the U.S.~Department of Energy.  Accordingly,
the U.S. Government retains a non-exclusive, royalty-free license to
publish or reproduce the published form of this contribution, or allow
others to do so, for U.S.~Government purposes.

\end{document}